\documentclass[prl,twocolumn,showpacs,floatfix]{revtex4}
\usepackage{graphics,epsfig,amsfonts,amssymb,amsmath}
\begin{document}
\title{Huge enhancement of the magnetoresistance and negative
  differential conductance in nanostructure arrays}
\author{V. Est\'evez}
\author{E. Bascones}
\affiliation{Instituto de Ciencia de Materiales de Madrid,
ICMM-CSIC, Cantoblanco, E-28049 Madrid (Spain).}
%\maketitle
\email{vestevez@icmm.csic.es,leni@icmm.csic.es}
%\date{\today}
%\title[\texttt{achemso}]{Huge enhancement of the magnetoresistance in
%  nanostructure arrays}
%\begin{document}
\begin{abstract}
We show that the interplay between charging effects and the non-equilibrium
spin accumulation has a dramatic effect in the current through an array of
nanostructures attached to ferromagnetic electrodes. Large oscillations in the
current as a function of bias voltage show up for parallel orientation 
of the electrodes' magnetizations.
These oscillations originate in the inhomogeneity of the spin potentials
through the array and  
correlate with oscillations in the spin
accumulation. For antiparallel orientation the spin potential is homogeneous
and the oscillations do not show up.
This sensitivity results in a huge enhancement of the tunneling
magnetoresistance as compared to the single-island case, 
and 
%Our results 
open new routes for improving the spintronic response
of nanodevices. 
\end{abstract}
%\pacs{75.10.Jm, 75.10.Lp, 75.30.Ds}
\maketitle

A lot of effort has been devoted in the last two decades to understanding and
controlling the interplay between magnetism and charge current
because of promising applications in electronic devices\cite{reviewawschalom,
  reviewjohnson, reviewfert,spintronics}.  
Interest is focused on systems with a large magnetoresistance (MR), defined in
terms of the resistance of the device for parallel (P) and antiparallel (AP) 
orientation of two ferromagnetic electrodes, $MR=(R_{AP}-R_{P})/R_P$. 
The tunneling magnetoresistance (TMR) of  magnetic tunnel junctions,
early studied by Julli\'ere\cite{julliere}, 
is controlled by the spin polarization of the
carriers $p$, assumed here to be equal in both electrodes:
\begin{equation}
TMR^{Jul}=\frac{2 p^2}{1-p^2}
\label{eq:julliere}
\end{equation}

In a magnetic double tunnel junction, with a metallic area inserted
between the ferromagnetic electrodes, the TMR vanishes except if the spin 
flip processes are of little importance in the metal. Otherwise, the 
tunneling electrons lose memory of their initial spin. When the spin 
relaxation time is long, spin accumulation happens  and induces a spin
splitting of the  chemical potential in the antiparallel 
configuration\cite{johnsonprl}. 
The TMR is finite but only the half of Julliere's value.
A further reduction of the magnetoresistance occurs with increasing number of
metallic non-magnetic insertions, $N$, separated among themselves by 
tunnel barries:  
\begin{equation}
TMR^N=\frac{TMR^{Jul}}{2\left (1+\frac{(1-p^2)(N-1)^2}{4}\right )}
\label{eq:tmrn}
\end{equation}   
where all the tunnel barriers separating the metallic insertions are supposed
to be equivalent. 
The reduction of the TMR with the number of metallic regions is expected as it
is only via the spin accumulation that the information about the relative
magnetic orientation of the electrodes is transferred. This mechanism is
indirect and less effective than the direct coupling of the two electrodes in
a single tunnel junction\cite{weymann2005}.

If the dimensions of the metallic insertion are reduced charging effects can
no longer be neglected and new phenomena appear in the current\cite{cb} and in 
the magnetoresistance\cite{ono97}. Due to the cost in energy to add an electron 
to the islands, at zero temperature, the current is blocked below a threshold 
voltage $V_{th}$. For a symmetrically biased array with 
nonmagnetic electrodes and short-range interactions 
$V_{th}\sim 2 N E_c$\cite{nosotrosprb08} . 
Above this threshold the current is non-linear in voltage and frequently shows
Coulomb staircase features.  For a single island, a finite
polarization of the electrodes can induce regions of negative differential 
conductance, as well as oscillations and changes of sign in the 
TMR, dependending on the specific 
set-up \cite{barnas98,maekawa98,barnasepl,brataas99,imamura99,yakushiji2005}. 
Charging effects are known to be enhanced in nanostructure 
arrays\cite{cb,likharev89,middleton93,nosotrosprb08}, 
however little is known about the spin transport in these systems. 
As the spin accumulation is the only way to transfer the spin information, a
reduction of the TMR with increasing number of islands, similar to that in 
Eq.(2) could be expected, a priori.

Here we show that, increasing the number of islands in between two 
ferromagnetic electrodes has a dramatic impact on the charge and spin 
transport. Large current oscillations show up in the I-V curves for 
parallel magnetic orientation, resulting in a strongly voltage dependent 
tunneling magnetoresistance. This behavior is a consequence of the interplay 
between charging effects and
the spin accumulation through the array. The latter is 
homogenous along the array in the antiparallel configuration while it is 
inhomogenous and changes sign for parallel orientation. 
Opposite to what happens in the absence of
charging effects, the TMR can be orders of magnitude 
larger than in the single particle case. 
 
%Our results show that
%a proper patterning can largely improve the response of magneto-devices.

We consider an array of $N$ metallic nanostructures, in the following 
called islands, with  charging energy $E_c$ and single particle level
spacing $\delta$ satisfying $\delta \ll  K_BT \ll E_c$. 
$T$ is the temperature and $K_B$ the Boltzmann constant. 
To allow for spin accumulation $\delta$ is kept finite.
The array is placed in between two ferromagnetic electrodes with spin 
polarization $p$. The islands are separated between themselves and from the
ferromagnetic electrodes by equivalent tunnel junctions. The electronic 
interactions are assumed to be finite only when the charges are in the same 
conductor i.e. capacitive coupling between 
different conductors vanishes, and charge disorder is absent.  
The electronic charge is taken equal to unity.
Transport is treated at the sequential tunneling level with tunneling rates
\begin{equation}
\Gamma_\sigma(\Delta E_\sigma)=\frac{1}{R_\sigma}\frac{\Delta
  E_\sigma}{exp(\Delta E_\sigma/K_BT)-1}
\end{equation} 
Here $\Delta E_\sigma$ is the change in energy of electrons with spin $\sigma$ 
due to the tunneling process. The ferromagnetic polarization of the electrodes
enters via the tunneling resistance. $R_\sigma$ is equal to $2 R_T$ at the internal 
junctions separating two 
islands and to $2 R_T(1\pm p)^{-1}$ at the contact junctions between an 
electrode 
and the neighboring island.  Plus (minor) signs are assigned to majority 
(minority) spin carriers. The spin is conserved in the tunneling and
in between tunneling
events. Magneto-Coulomb\cite{ono97,vandermolen06,kouwenhoven09}, 
spin-orbit effects, and the role of the metal insulator 
interface\cite{deteresa99} in determining the 
spin polarization of the carriers are neglected through all the paper.

Under these assumptions the spin dependent potential at island $i$ is
$\phi_{i,\sigma}=N_{i,\sigma} \delta + 2 (N_{i,\sigma} + N_{i,-\sigma})E_c$
with $i$ running from $1$ to $N$ and $N_{i,\sigma}$ the number of excess 
electrons with spin $\sigma$ at island $i$. 
Correspondingly the spin accumulation at island $i$ is given by  
$\Delta \phi _{i,\sigma}=(N_{i,\sigma}-N_{i,-\sigma}) \delta$.   
The electrodes are maintained at spin-independent potentials 
$\phi_{0,\sigma}=V/2$ and $\phi_{N+1,\sigma}=-V/2$ and current flows 
perpendicular to the array axis.

To calculate the current we use a Monte Carlo simulation which depends on the
tunneling rates, as described in previous works\cite{likharev89,nosotrosprb08}. 
At each iteration a single tunneling event takes place. The time involved in 
this event depends on the  tunneling rates of all the possible tunneling
processes. These rates are calculated at each iteration as they change 
when the spin and charge state of the system does.  
The values of the current given here correspond to stationary states. 
This computational method is appropriate for low temperatures when  the
electrons flow in the same direction in all 
the relevant tunneling processes. 
With increasing temperature the electrons become frequently stacked, 
oscillating back and forth between two islands and computation is not 
possible.  
Calculations reported here correspond to $p=0.7$, $\delta=10^{-5}$ and 
$K_BT=10^{-4}$ with energies given through all the paper in units of
$E_c$. The value of $\delta$ does not affect the results as far as $\delta
\ll K_BT$ is satisfied. The non-zero temperature used to ensure 
$\delta \ll K_BT$  
allows a very small but finite current below threshold values. For simplicity, 
in most part of the discussion this finite temperature effect is
neglected. 

\begin{figure}
\leavevmode
\includegraphics[clip,width=0.5\textwidth]{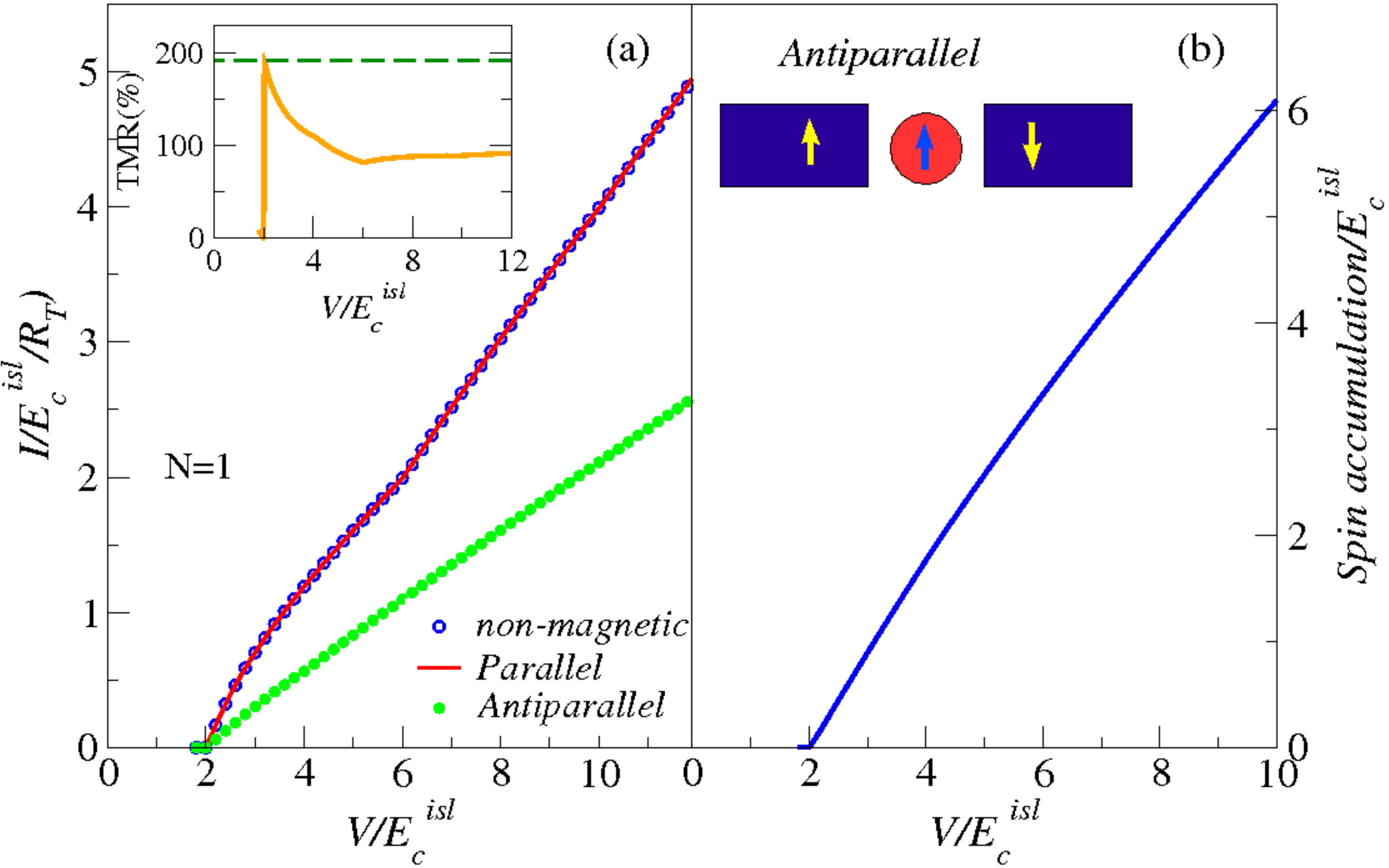}
\caption{(a) Main figure: I-V curves of a single island placed in between two
  metallic (non-magnetic) or ferromagnetic electrodes for $p=0.7$ with 
  parallel and
  antiparallel orientation of the electrode magnetizations, see text. 
  The corresponding TMR is shown  as a function of 
  voltage  in the inset (a) where the Julliere's TMR is marked with a dashed
  line. (b) Main figure: Spin accumulation  for a
  single island and antiparallel arrangement. 
  Inset in (b): Sketch of the device and the spin accumulation for
  antiparallel orientation. The arrow in the island shows the spin
  with largest potential.} 
\label{fig:oneisland}
\end{figure}

Results for a single island are shown in Fig. 1. For parallel arrangement 
the current can be described in terms of
two independent spin channels and has the same value found with nonmagnetic
electrodes, see Fig. 1a. For antiparallel orientation  spin
accumulation appears to equalize the ratios for entrance and exit from the 
island of spin up and down electrons, see inset in Fig. 1b. Both the current 
and the spin accumulation increase monotonously with voltage above $V_{th}$ as
shown in Fig.~1. 
The inset in Fig. 1a shows the TMR corresponding to a single island as a 
function of the
bias voltage $V$. When $V \gg E_c$ charging effects are not important 
and the TMR equals half of Julliere's value, as expected for a double tunnel 
junction. As the voltage is reduced the TMR first decreases slightly and 
then increases until $V_{th}$ where it suddenly jumps to zero. 
The maximum TMR, found at $V_{th}$ equals Julliere's value given in 
Eq. (\ref{eq:julliere}).

A very different behavior is found for $N \geq 3$, as the inner junctions
between the islands can block the current flow. In the antiparallel
configuration, as for non-magnetic electrodes, the current is strongly 
suppressed below $V\sim V_{th}$ where a large jump is observed, see
Figure 2(a). Surprisingly for parallel orientation of the electrode 
magnetization the current show large 
oscillations below $V_{th}$.  Within the range of voltages at which 
these oscillations are observed, there are regions in which 
the current decreases with increasing $V$, corresponding to negative
differential conductance. This negative conductance can be quite large, see 
the inset of Fig. 2 (a). As observed in Figure 2(b), 
the number of peaks observable in the current increases with the number of 
islands in the array.

This unexpected behavior can be understood from an analysis of the potential
drop through the array.
In the case of metallic non-magnetic electrodes, the large threshold voltage originates in
the lack of potential drop at the inner junctions which impede the flow of
charges below $V_{th}$. This situation is typical of short-range interactions,
as the voltage drops mostly at the junctions adjacent to source and 
drain, see discussion in \cite{nosotrosprb08}. With increasing voltage charges are allowed to enter the array 
from the electrodes and create a charge gradient which provides the
potential drops at the inner junctions 
necessary to allow the tunneling processes. 
Once a single charge occupies the first island, to increase the
number of charges on it a voltage rise equal to $4E_c$ is needed.   
$V_{th}$ is the voltage necessary to put around $N/2$ charges onto the 
first island (the exact number depends on $N$ being even or odd). 

\begin{figure}
\leavevmode
\includegraphics[clip,width=0.5\textwidth]{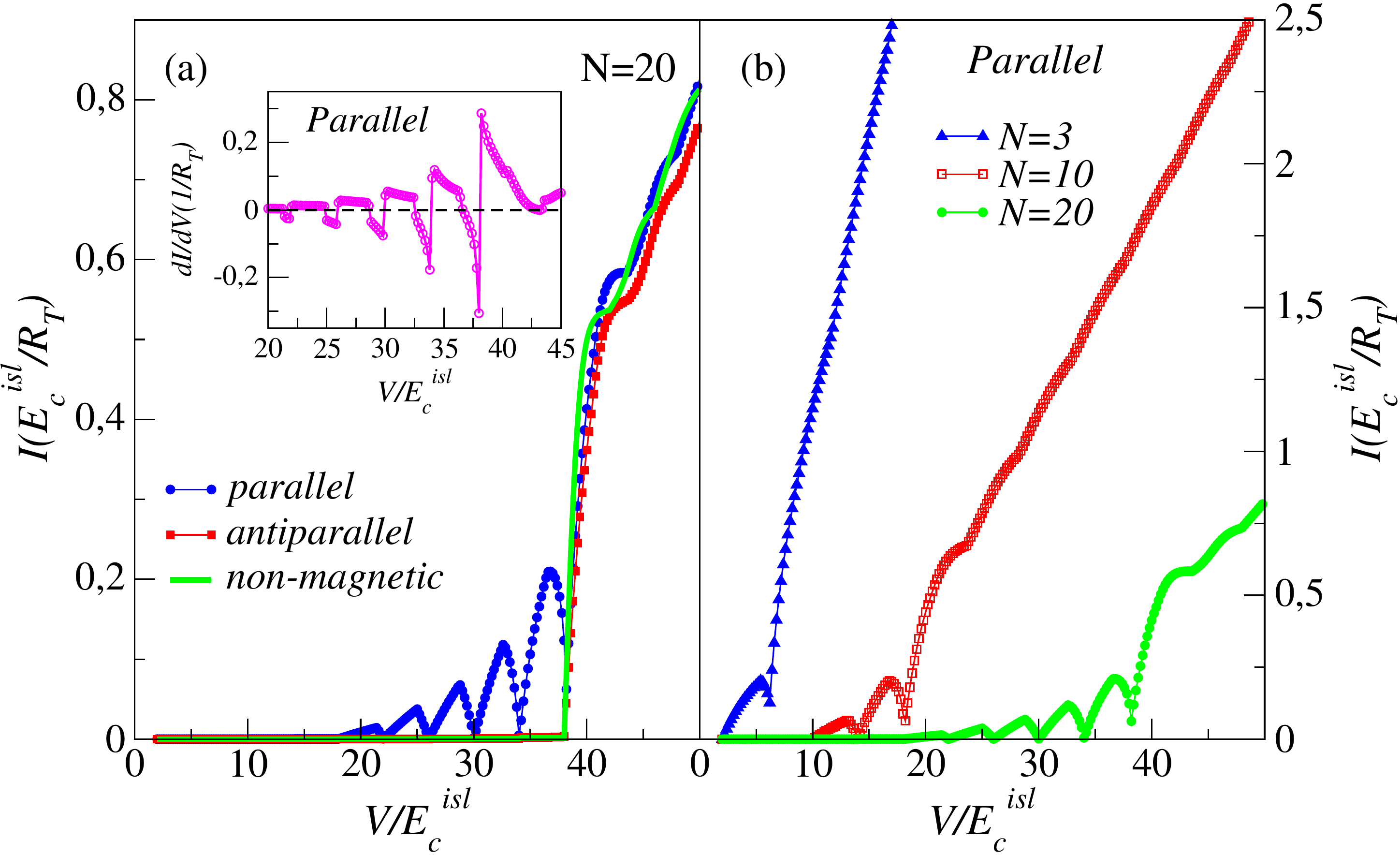}
\caption{(a) I-V curves for an $N=20$ array with ferromagnetic 
electrodes in parallel and antiparallel arrangement. The I-V curve
corresponding to nonmagnetic electrodes is plotted to serve as a
reference. Inset: Differential conductance corresponding to $N=20$ and
parallel arrangement. (b) I-V curves for arrays of different sizes and parallel
orientation of the electrode magnetizations. The number of peaks in the I-V
curve increases with the size of the array.} 
\label{fig:ivarray}
\end{figure}

Because of the spin accumulation when the electrodes are magnetic the
potential profile is spin dependent. In the antiparallel configuration the 
spin-up and spin-down potentials change only very slightly when moving 
from an island to its neighboring ones,
see Fig. 3(a). 
The spin accumulation is very homogeneous through the array. So, it
barely contributes to the change in energy for tunneling at the inner
junctions. As for non magnetic electrodes, this change in energy is controlled
by the charge gradient at the junctions. The position dependence of the spin accumulation 
is very different in the antiparallel and
parallel cases, see the sketches given at the top of Figure 3.  
For parallel orientation of the electrodes magnetization the spin accumulation 
changes sign as one moves from the
first island to the end of the array, as seen in 
Fig. 3(b), where we plot the spin potentials 
corresponding to $N=20$ and $V=20 E_c$ as a function of the
island position.  Spin down  (up) potential 
increases (decreases) producing potential drops at the junctions between 
the islands which oppose (favor) the current flow. Because of this potential 
drop created by the spin accumulation the charge gradient needed to allow the 
flow of spin up electrons decreases. With the reduction of the number of charges
which have to be accumulated at the first or last island to allow current, the
threshold voltage is reduced. Current is observed at smaller voltages for 
parallel arrangement of the electrode magnetizations than in the antiparallel
or nonmagnetic cases. 
With the potential drop at the inner junctions being spin dependent, 
naively one could expect a complete spin polarization of
the current. As observed in Fig. 3(c), this does not happen. The current
polarization  is not complete because  
the flow of spin-down electrons is allowed via the
charge potential left by the spin-up ones. 

\begin{figure}
\leavevmode
\includegraphics[clip,width=0.5\textwidth]{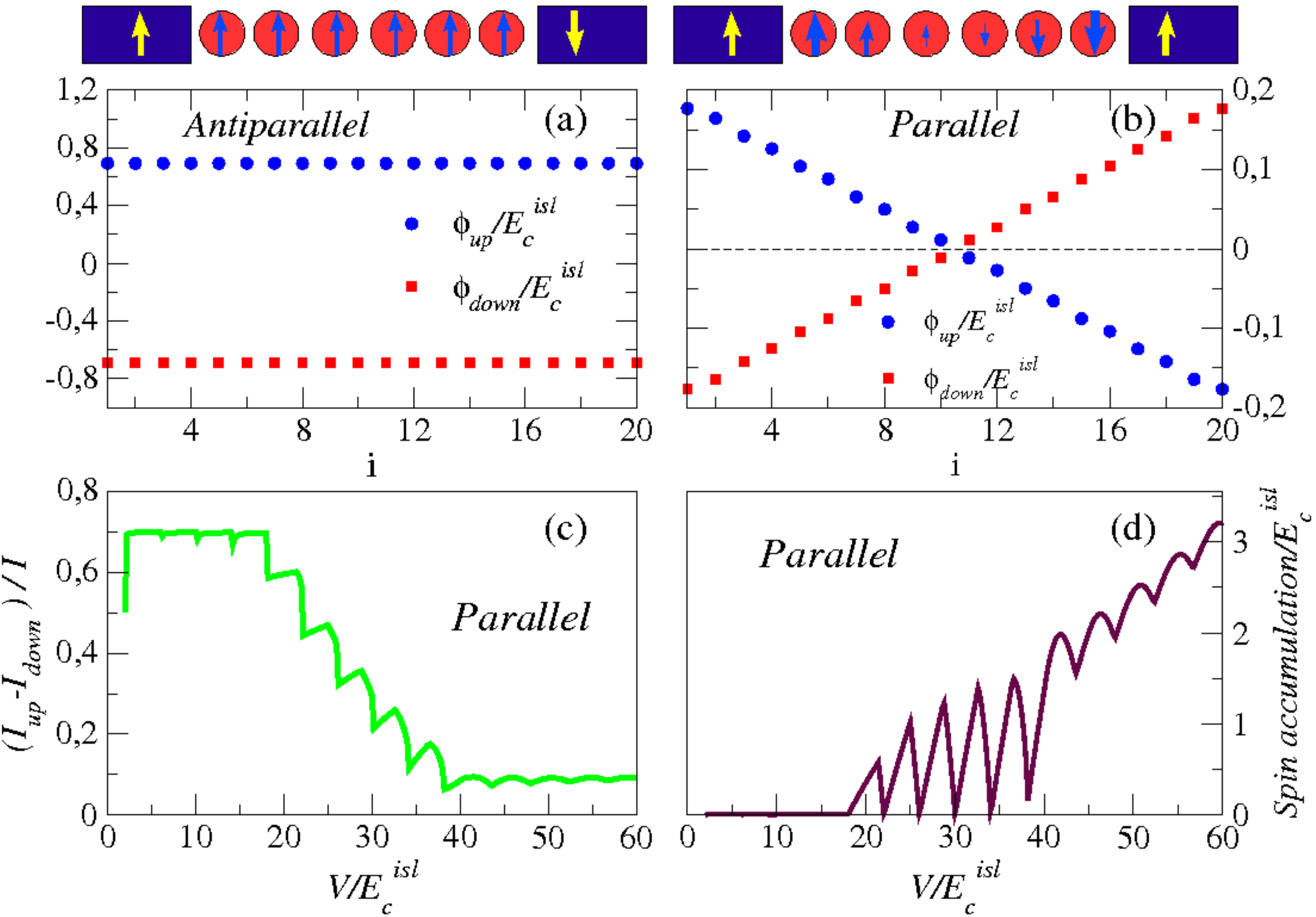}
\caption{Top: Sketch of the spin accumulation in an N=6 array with 
antiparallel (left) and
  parallel configuration. Middle: Spin potentials as a function of island position at $V=20 E_c$ for an
  $N=20$ array  corresponding to antiparallel (a) and parallel (b) 
 orientation of the magnetization electrodes. Bottom: (c) and (d) show the voltage 
dependent spin polarization of the current and the spin accumulation in the 
first island for the same $N=20$ array and parallel orientation.}
\label{fig:spinpotential}
\end{figure}

If, as discussed above, the current flows thanks to the spin accumulation, 
one could expect
a decrease in the latter correlated with the regions of negative differential 
conductance. This is confirmed in Fig. 3(d) 
which shows oscillations in the spin
accumulation at the same voltages as in the current. The decrease in the spin
accumulation is due to the opening of a new transport channel only for spin down 
electrons. This is possible because the voltage drop
necessary to permit the entrance of spin-down charges at the first island is
smaller than the one corresponding to spin-up ones. At some $V$, being the first 
island occupied with $n$ charges, spin-down electrons would be allowed to
enter, while spin-up electrons will not. As spin accumulation happens to
equilibrate the ratio between entrance and exit of spin up and down electrons,
the opening of a new conduction channel for down electrons will reduce the
spin accumulation. This is the origin of the decrease of current with
increasing voltage, which lasts until this conduction channel is also 
opened for spin-up electrons. 

\begin{figure}
\leavevmode
\includegraphics[clip,width=0.5\textwidth]{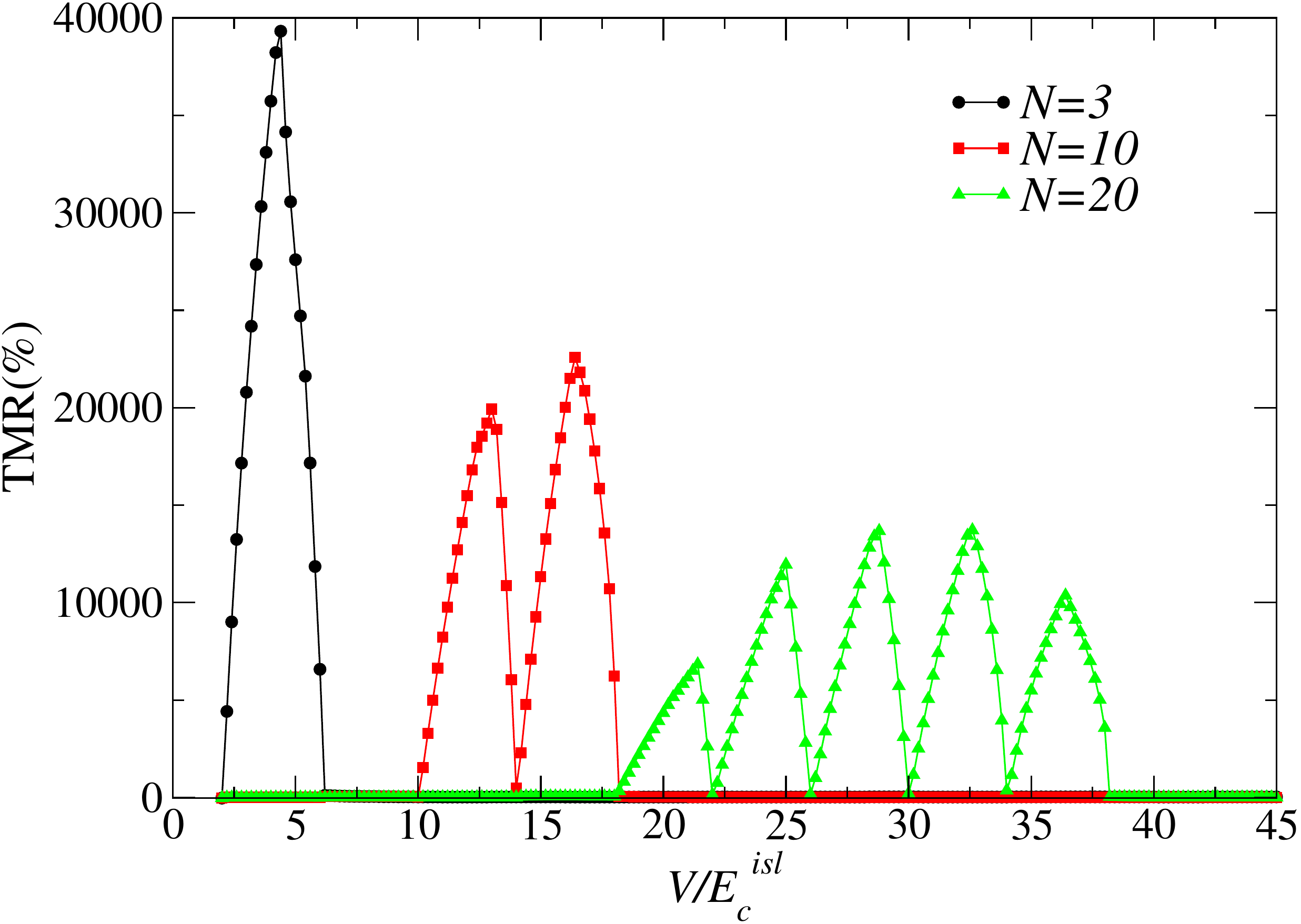}
\caption{Tunneling magnetoresistance as a function of bias voltage for
  different array sizes.}
\label{fig:tmr}
\end{figure}
     
The differences in the current as a function of a magnetic orientation have a
dramatic effect on the TMR, plotted in Figure 4, which is strongly
voltage dependent. Like the current it shows peaks in the parallel configuration. But the most impressive result are the extremely large values of the 
TMR observed. Even more than two orders of magnitude larger than 
in the single island case and than the value predicted by Eq. (2) 
which is recovered at large
voltages (not shown). The largest values of the magnetoresistance depend on
voltage and are largest for $N=3$. This shows that there is a lot of room to
improve the TMR response of electronic devices.  

In summary, we have studied the interplay between magnetism and charging
effects in the transport through nanostructure arrays placed between
ferromagnetic electrodes. We have found very
non-linear I-V curves for parallel orientation of source and drain
magnetizations, which include peaks in the current at voltages smaller 
than the metallic
threshold voltage. This unusual 
dependence originates in the inhomogeneity of the spin
potential through the array. 
The oscillations of the current correlate with
oscillations of the TMR and of the 
spin accumulation which changes sign through the array.
The TMR can reach values two orders of magnitude larger than those found in
the single island case. 
Our results show that in the presence of charging effects 
a proper patterning can largely enhance the value of the magnetoresistance 
and will stimulate further experimental and theoretical studies.  
   
We are thankful to J. K\"onig for proposing us this analysis and 
for enlightening discussions. V. Est\'evez thanks the hospitality of the 
University of Duisburg-Essen where this work was initiated.  
Funding from Ministerio de Ciencia e Innovaci\'on through
Grants No. FIS2008-00124/FIS, FPI fellowship and Ram\'on y Cajal contract, and
from Consejer\'ia de Educaci\'on de la Comunidad Aut\'onoma de Madrid and CSIC
through Grants No. CCG07-CSIC/ESP-2323, CCG08-CSIC/ESP3518, PIE-200960I033 is 
acknowledged.

\end{document}